\documentclass[preprints,article,accept,moreauthors,pdftex]{mdpi} 

\usepackage{amssymb}
\usepackage{color}
\usepackage{subfigure}
\usepackage{bm}
\usepackage{physics}
\usepackage{aas_macros}
\usepackage{wrapfig}
\usepackage{gensymb}

\firstpage{1} 
\pubvolume{xx}
\issuenum{1}
\articlenumber{5}
\pubyear{2021}
\copyrightyear{2021}
\history{Received: date; Accepted: date; Published: date}

\pdfoutput=1

\Title{Timescales for Detecting Magnetized White Dwarfs in Gravitational Wave Astronomy}

\Author{Surajit Kalita \orcidA{}}

\AuthorNames{Surajit Kalita}

\address{%
\quad Department of Physics, Indian Institute of Science, Bangalore 560012, India; surajitk@iisc.ac.in}

\conference{} 

\abstract{Over the past couple of decades, researchers have predicted more than a dozen super-Chandrasekhar white dwarfs from the detections of over-luminous type Ia supernovae. It turns out that magnetic fields and rotation can explain such massive white dwarfs. If these rotating magnetized white dwarfs follow specific conditions, they can efficiently emit continuous gravitational waves and various futuristic detectors, viz. LISA, BBO, DECIGO, and ALIA can detect such gravitational waves with a significant signal-to-noise ratio. Moreover, we discuss various timescales over which these white dwarfs can emit dipole and quadrupole radiations and show that in the future, the gravitational wave detectors can directly detect the super-Chandrasekhar white dwarfs depending on the magnetic field geometry and its strength.}

\keyword{White dwarfs; gravitational waves; magnetic fields; rotation; luminosity}  

\begin{document}

\section{Introduction}\label{Introduction}

Over the years, the luminosities of type Ia supernovae (SNeIa) are used as one of the standard candles in cosmology to measure distances of various astronomical objects. This is due to the reason that the peak luminosities of all the SNeIa are similar, as they are originated from the white dwarfs (WDs) burst around the Chandrasekhar mass-limit ($\sim 1.4M_\odot$ for carbon-oxygen non-rotating non-magnetized WDs \cite{1931ApJ....74...81C}). However, the detection of various over-luminous SNeIa, such as SN 2003fg \cite{2006Natur.443..308H}, SN 2007if \cite{2010ApJ...713.1073S}, etc., for about a couple of decades, questions the complete validity of the standard candle using SNeIa. Calculations show that such over-luminous SNeIa needs to be originated from WDs with masses $\sim 2.1-2.8 M_\odot$. It is well understood that these WDs exceed the standard Chandrasekhar mass-limit, and hence, they are termed as the super-Chandrasekhar WDs. Ostriker and Hartwick first showed that rotation could increase a WD's mass up-to $\sim 1.8M_\odot$ \cite{1968ApJ...153..797O}. Later, Mukhopadhyay and his collaborators, in a series of papers for about a decade, showed that magnetic field could be one of the prominent physics, which can significantly increase the mass of the WDs \cite{2012PhRvD..86d2001D,2015MNRAS.454..752S,2019MNRAS.490.2692K,2020IAUS..357...79K}. Nevertheless, various other physics such as modified gravity \cite{2018JCAP...09..007K}, generalized uncertainty principle \cite{2018JCAP...09..015O}, noncommutative geometry \cite{2021IJMPD..3050034K}, and many more, can also explain these super-Chandrasekhar WD progenitors.

So far, no super-Chandrasekhar WDs have been observed directly in any of the surveys such as GAIA, SDSS, Kepler, etc. as the highly magnetized WDs generally possess very low thermal luminosities \cite{2020MNRAS.496..894G}. Hence, they are difficult to be observed in the standard electromagnetic (EM) surveys. On the other hand, the detection of gravitational waves (GW) by LIGO/Virgo from the various compact object merger events opens a new era in astronomy. In the future, space-based GW detectors such as LISA, DECIGO, BBO, etc., can detect many other astrophysical objects. In this article, we show that these space-based detectors can also detect the continuous GW emitted from the magnetized super-Chandrasekhar WDs, provided they behave like pulsars. We also show that the timescale for the emission of the dipole and quadrupole radiations from these WDs is highly dependent on the magnetic field geometry and its strength.

\section{Structure of rotating magnetized white dwarfs}
\begin{wrapfigure}{r}{0.3\textwidth}
	\begin{center}
		\includegraphics[width=0.33\textwidth]{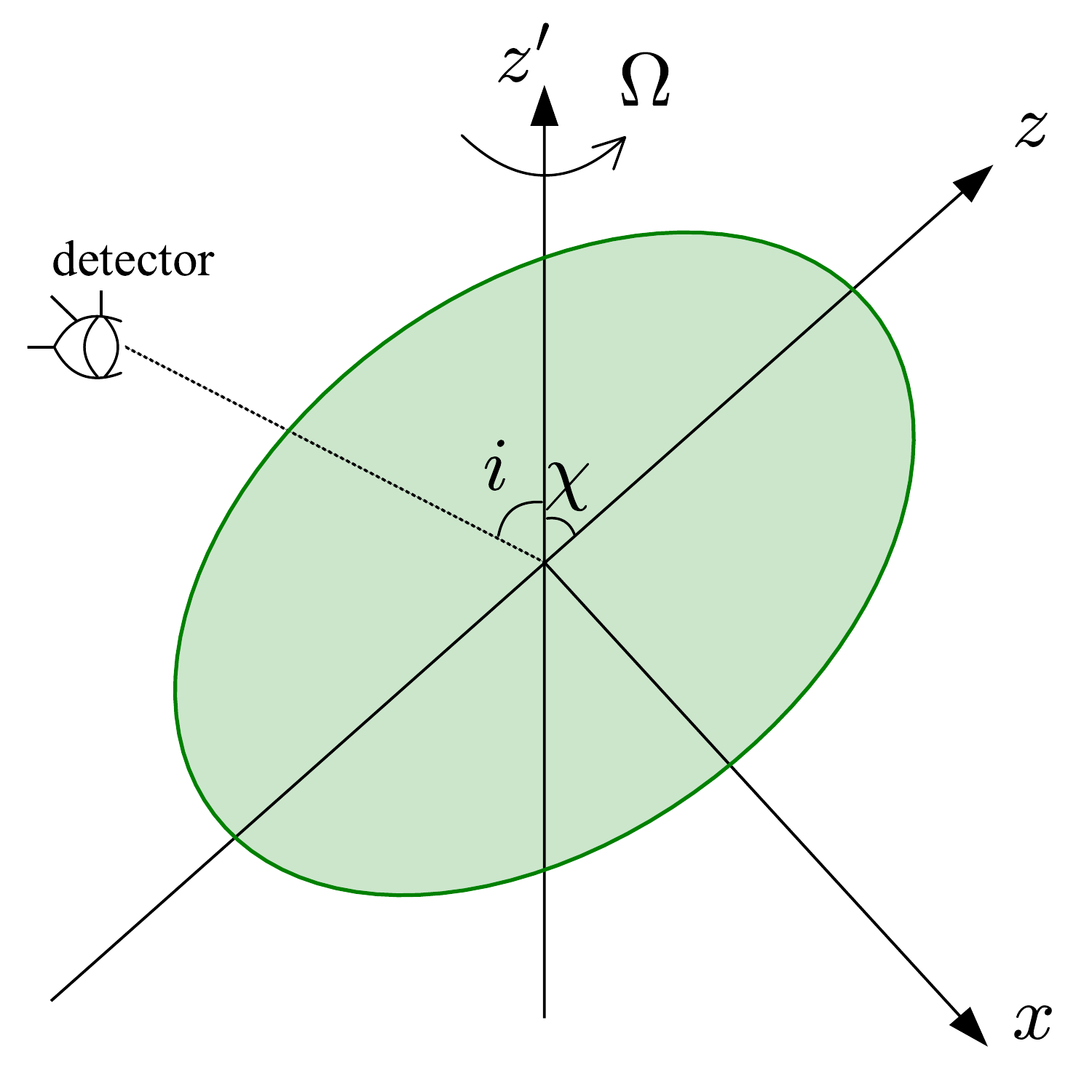}
	\end{center}
	\caption{A cartoon diagram of magnetized rotating WD pulsar. Here $z'$ and $z$ are respectively the rotation and magnetic field axes.}
	\label{Fig: Compact objects}
\end{wrapfigure}
\begin{figure}[htbp]
	\centering
	\subfigure[Toroidal magnetic field]{\includegraphics[scale=0.45]{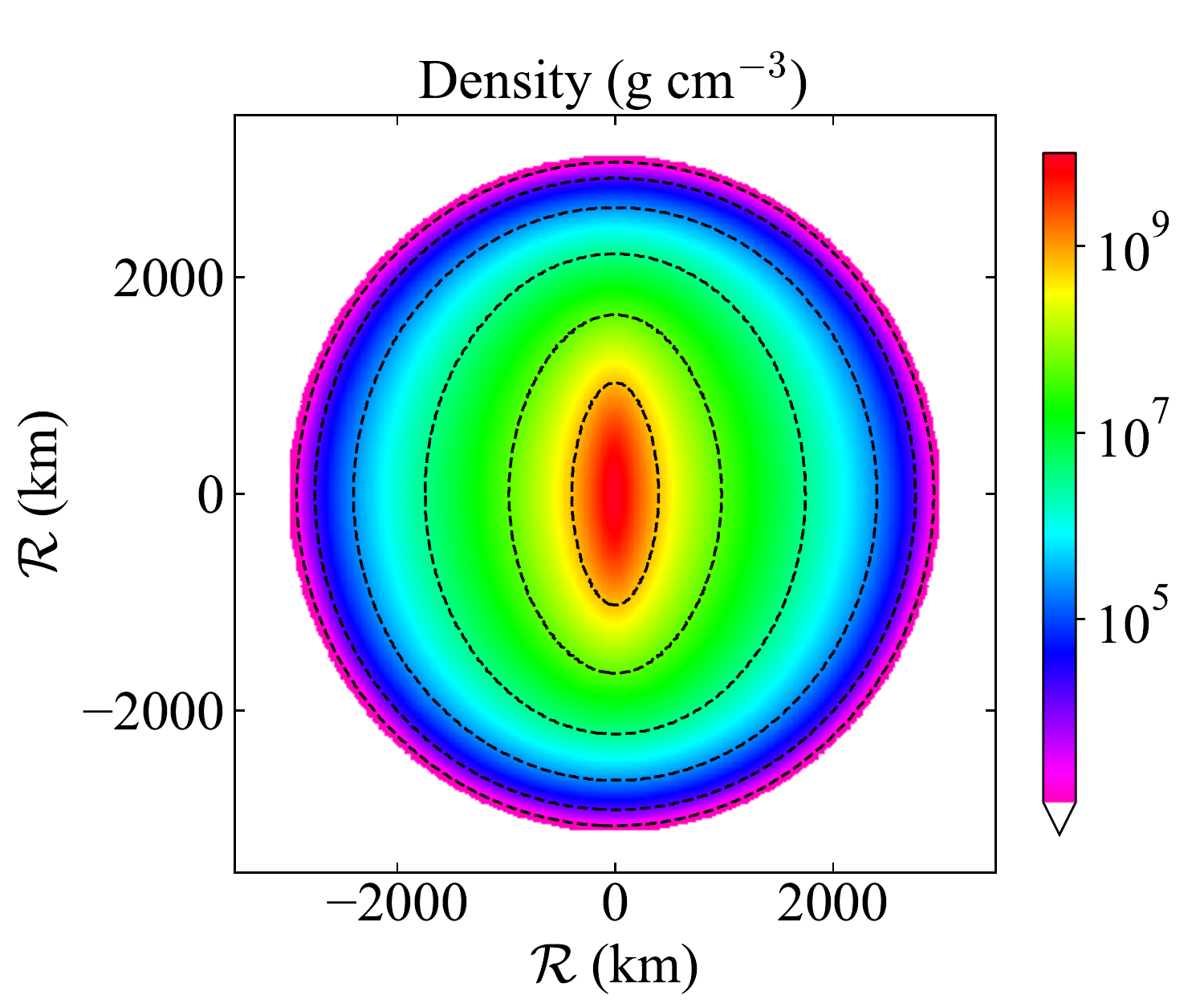}}
	\subfigure[Poloidal magnetic field]{\includegraphics[scale=0.45]{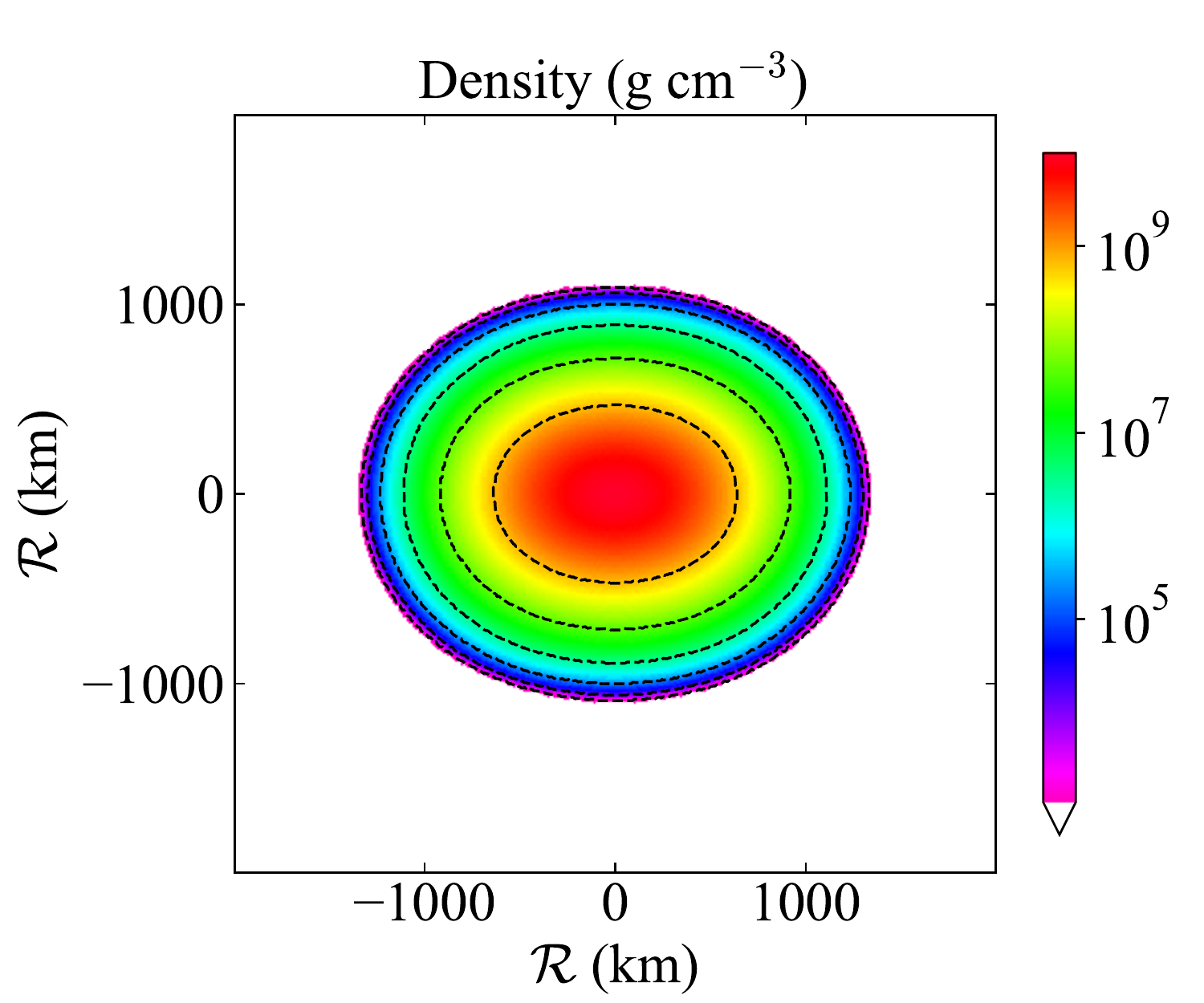}}
	\caption{Density isocontours of magnetized WDs with central density $\rho_c=10^{10}$ g cm$^{-3}$.}
	\label{Fig: Magnetized WD}
\end{figure}

A system can emit GW if it possesses a time-varying quadrupolar moment. Hence, neither a spherically symmetric nor an axially symmetric WD can emit GW. It is well known that a toroidal field can make a star prolate-shaped, whereas rotation and poloidal magnetic fields can transform it into oblate-shaped. Hence, it is necessary to have a misalignment between the magnetic field and rotation axes to make the WD a tri-axial system so that it can emit gravitational radiation. For a WD, as shown in Figure \ref{Fig: Compact objects}, rotating with an angular frequency $\Omega$,  the two polarizations of GW, $h_+$ and $h_\times$, at a time $t$ are given by \cite{1979PhRvD..20..351Z,1996A&A...312..675B}
\begin{equation}\label{gravitational polarization}
\begin{aligned}
h_+ &= h_0\sin\chi\left[\frac{1}{4}\sin 2i\cos\chi\cos\Omega t-\frac{1+\cos^2i}{2}\sin\chi\cos2\Omega t\right],\\
h_\times &= h_0\sin\chi\left[\frac{1}{2}\sin i\cos\chi\sin\Omega t-\cos i\sin\chi\sin2\Omega t\right],
\end{aligned}
\end{equation}
with
\begin{equation}\label{grav_wave_amplitude}
h_0 = \frac{4G}{c^4}\frac{\Omega^2\epsilon I_{xx}}{d},
\end{equation}
where $\chi$ is the angle between the rotation and magnetic field axes, $i$ is the angle between the rotation axis and the observers' line of sight, $G$ is Newton's gravitational constant, $c$ is the speed of light, $d$ is the distance between the detector and the source, and $\epsilon = |I_{zz}-I_{xx}|/I_{xx}$ with $I_{xx}$ and $I_{zz}$ being the moments of inertia of the WD about $x-$ and $z-$ axes respectively. It is evident that if $\chi$ is close to $0\degree$, radiation at frequency $\Omega$ is dominated, whereas if $\chi$ is near $90\degree$, the same at frequency $2\Omega$ is dominant. To model the rotating magnetized WDs, we use the {\it XNS} code \cite{2014MNRAS.439.3541P}, which solves the axisymmetric equilibrium configuration of stellar structure in general relativity. A detailed discussion about all the parameters used to model the WDs is given by Kalita and Mukhopadhyay \cite{2019MNRAS.490.2692K}. Figure \ref{Fig: Magnetized WD} shows the density isocontours for the WDs with purely (a) toroidal and (b) poloidal magnetic field geometry generated with {\it XNS} code. It is evident that the toroidal field makes the WD bigger, whereas the poloidal field reduces its size. The stability of a WD in the presence of rotation and magnetic fields can be understood respectively from the constraints on kinetic-to-gravitational energy ratio \cite{1989MNRAS.237..355K}, and magnetic-to-gravitational energy ratio \cite{2009MNRAS.397..763B}.

\section{Gravitational radiation emitting from rotating magnetized white dwarfs}

\begin{figure}[htpb]
	\centering
	\includegraphics[width=0.61\textwidth]{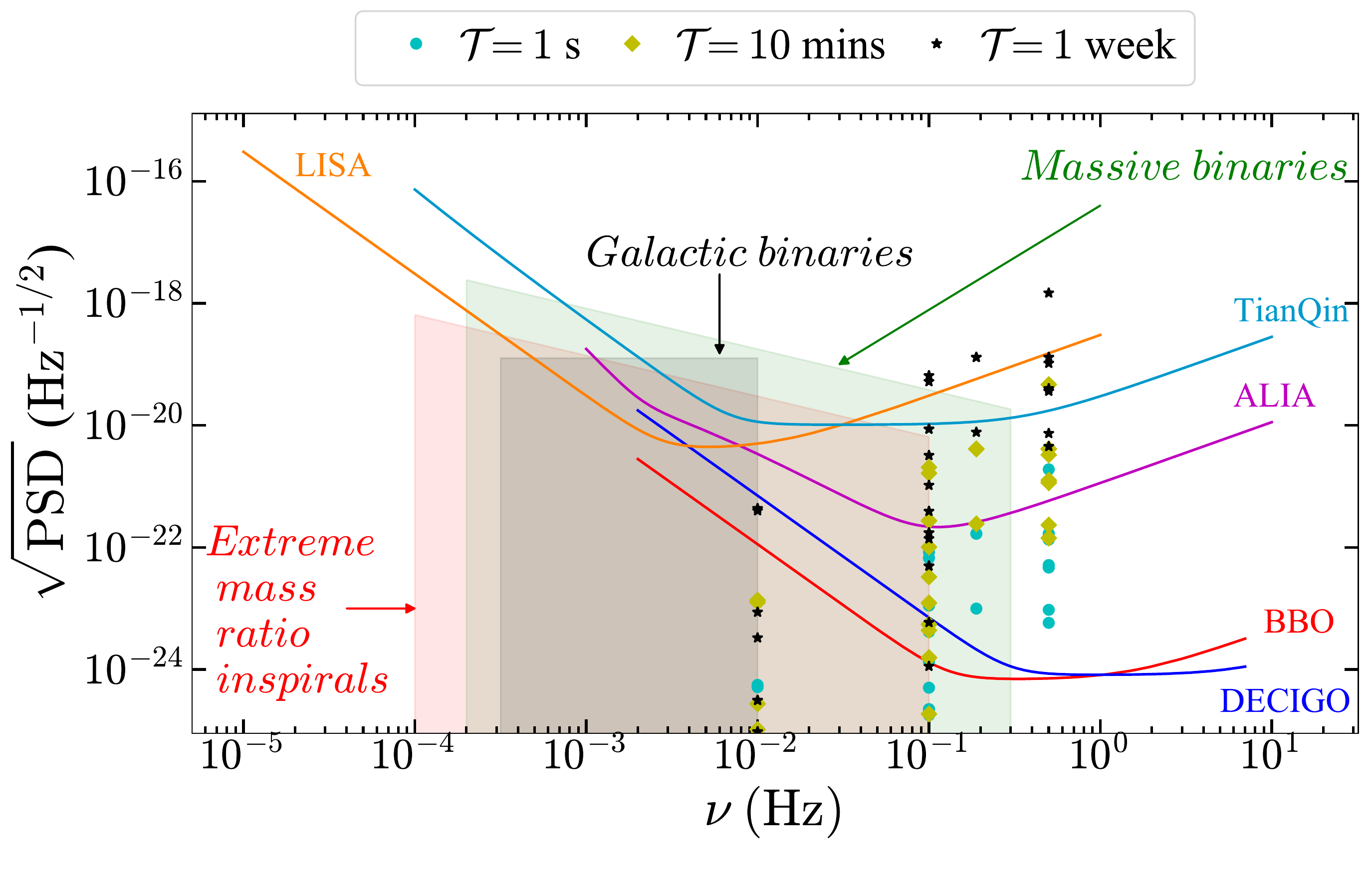}
	\caption{$\sqrt{\mathtt{S}(\nu)}$ of different GW detectors along with $\sqrt{\mathcal{T}/5}h$ for the various rotating magnetized WDs with different integration time. We assume $\chi=30\degree$ and $d=100$ pc.}
	\label{Fig: Detector_PSD}
\end{figure}
We model several rotating WDs using {\it XNS} code with different central densities considering various magnetic field geometry and strength. The relation between the integrated signal-to-noise ratio (SNR) and the observation time $\mathcal{T}$ is given by \cite{Maggiore}
\begin{equation}
\mathrm{SNR} = \frac{1}{\sqrt{5}}\sqrt{\frac{\mathcal{T}}{\mathtt{S}(\nu)}}h,
\end{equation}
where $\mathtt{S}(\nu)$ is the power spectral density (PSD) of the detector at a frequency $\nu$ and $h$ is the strength of GW at that frequency. Figure \ref{Fig: Detector_PSD} shows $\sqrt{\mathtt{S}(\nu)}$ for the various detectors \cite{2009LRR....12....2S,2020PhRvD.102f3021H} along with $\sqrt{\mathcal{T}/5}h$ for our modeled WDs. It is evident that BBO and DECIGO can immediately detect the magnetized WDs, whereas the minimum integrated time of observations for ALIA, TianQin and LISA turns out to be a few minutes, a few days and a few weeks, respectively. Moreover, the rotational frequencies for many of these WDs can be significantly higher than the orbital frequencies of galactic binaries, and hence, they are free from these confusion noise \cite{2010ApJ...717.1006R}. Other problems of confusion noise coming from the massive binaries can be sorted out with proper source modeling, which is beyond the scope of this work.

\section{Timescale for emission of gravitational radiation}

\begin{figure}[htbp]
	\centering
	\subfigure[$L_\mathrm{D} \gg L_\mathrm{GW}$]{\includegraphics[scale=0.47]{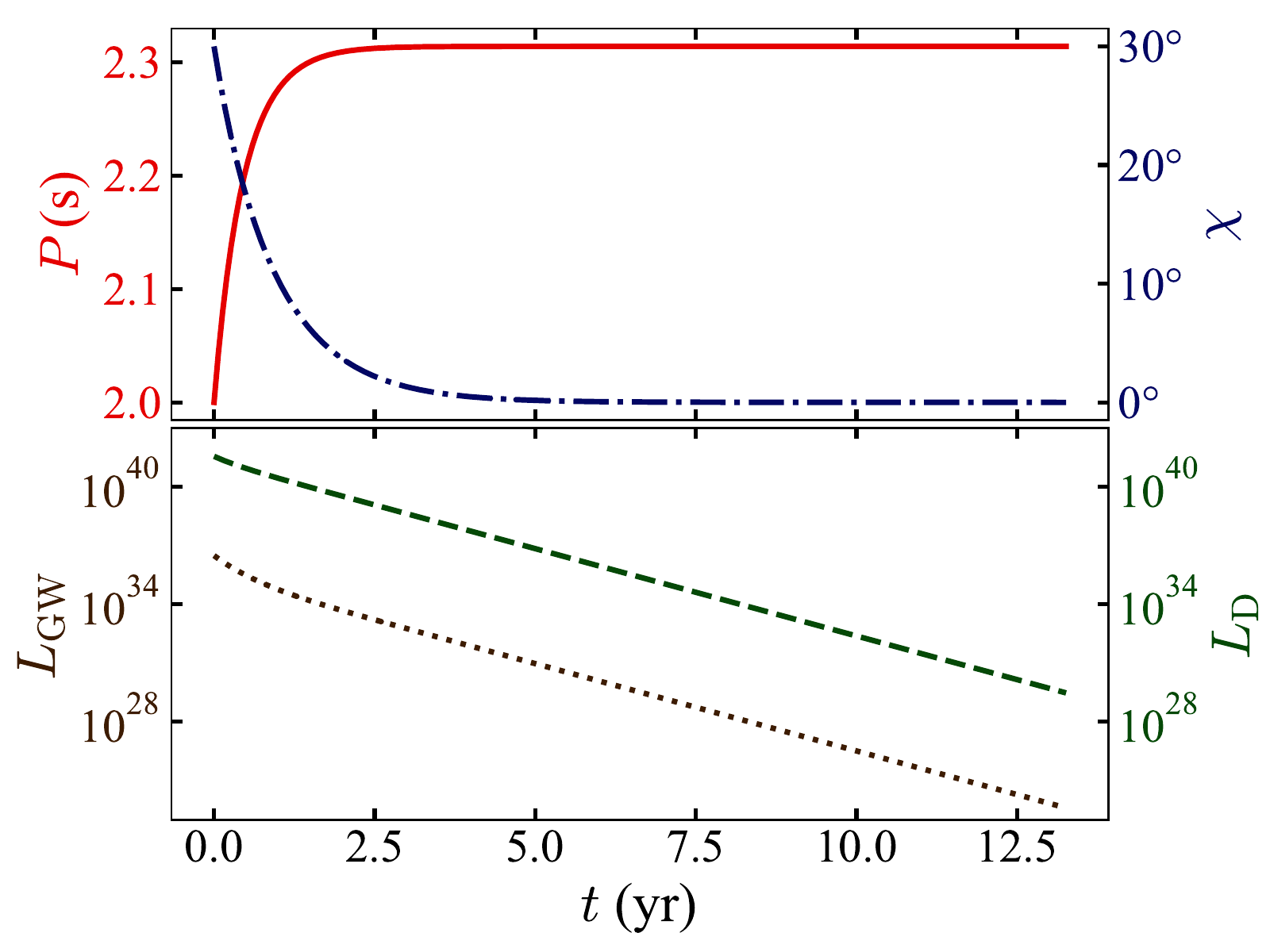}}
	\subfigure[$L_\mathrm{GW} \gg L_\mathrm{D}$]{\includegraphics[scale=0.47]{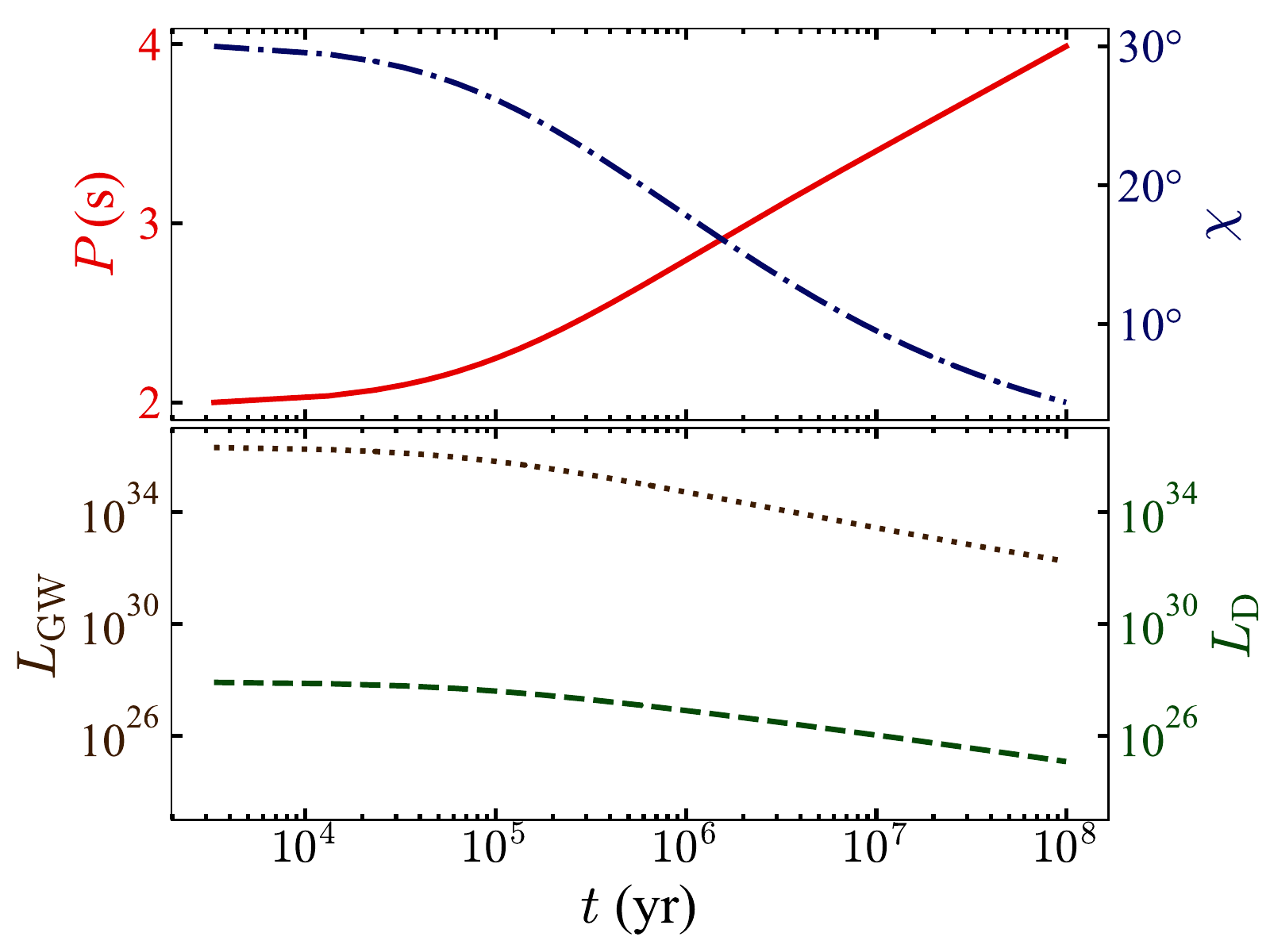}}
	\caption{Variation of $P$, $\chi$, $L_\mathrm{D}$, and $L_\mathrm{GW}$ with respect to time.}
	\label{Fig: Timescale}
\end{figure}

Since the WDs, we considered, as shown in Figure \ref{Fig: Compact objects} are pulsar-like, they can emit both dipole and quadrupole radiations, which are respectively associated with the dipole luminosity $L_\mathrm{D}$ and quadrupole luminosity $L_\mathrm{GW}$. Due to these radiations, the variations of $\Omega$ and $\chi$ with respect to $t$ are given by \cite{2000MNRAS.313..217M}
\begin{align}\label{Eq: omega}
\frac{d(\Omega I_{z'z'})}{dt} &= -\frac{2G}{5c^5} \left(I_{zz}-I_{xx}\right)^2 \Omega^5 \sin^2\chi \left(1+15\sin^2\chi\right)- \frac{2B_p^2 R_p^6 \Omega^3}{c^3}\sin^2\chi ~F(x_0),\\ \label{Eq: chi}
I_{z'z'} \frac{d\chi}{dt} &= -\frac{12G}{5c^5} \left(I_{zz}-I_{xx}\right)^2 \Omega^4 \sin^3\chi \cos\chi - \frac{2B_p^2 R_p^6 \Omega^2}{c^3}\sin\chi \cos\chi ~F(x_0),
\end{align}
where $I_{z'z'}$ is the moment of inertia of the WD along $z'-$axis, $x_0=R_0 \Omega/c$, $R_p$ is the radius of the pole, $R_0$ is the average radius, $B_p$ is the strength of the magnetic field at the pole of the WD, and the function $F(x_0)$ is defined as
\begin{equation}
F(x_0) = \frac{x_0^4}{5\left(x_0^6 - 3x_0^4 + 36\right)} + \frac{1}{3\left(x_0^2 + 1\right)}.
\end{equation}
Solving equations \eqref{Eq: omega} and \eqref{Eq: chi} simultaneously, we obtain the temporal variation of $\Omega$ and $\chi$. Both $\Omega$ and $\chi$ decreases with time, which leads to the decrease in $L_\mathrm{D}$ and $L_\mathrm{GW}$ as shown in Figure \ref{Fig: Timescale}. Spin period $P$ is defined as $P=2\pi/\Omega$. If $L_\mathrm{D} \gg L_\mathrm{GW}$, $\chi$ reduces to zero very quickly with a new spin period, and hence, the WD stops radiating soon. On the other hand, if $L_\mathrm{GW} \gg L_\mathrm{D}$, the WD can emit both dipole and quadrupole radiations for a long time. A detailed discussion with the exact expressions for these various cases is given by Kalita et al. \cite{2020ApJ...896...69K}.

\section{Conclusions}
We know that an object possessing a poloidal magnetic field can emit dipole radiation, whereas, with a toroidal field, no dipole emission is possible. However, in both the cases, quadrupole radiation is emitted as the WD structure is pulsar-like, which is a tri-axial system. If the WD has a strong poloidal field, it will have $L_\mathrm{D} \gg L_\mathrm{GW}$, and since in this case, $\chi$ quickly becomes zero, it stops radiating for a long time. On the other hand, in the case of a WD with a dominant toroidal field, $L_\mathrm{GW} \gg L_\mathrm{D}$ is satisfied, and it can radiate for a long time. Moreover, the birth rate of WD is $\sim 10^{-12}$ pc$^{-3}$ yr$^{-1}$ \cite{1983Ap&SS..97..305G}, which means that within 100 pc radius, on average, only one WD forms in $10^6$ yr, which is quite slow for a GW detector to detect. Hence, if the WD has a predominant toroidal magnetic field at the time of its birth, the GW detectors, such as LISA, BBO, DECIGO, and ALIA, can detect it for a long time. On the other hand, if the WD has a strong poloidal field, $h$ decreases quickly and so the SNR, and hence, it can hardly be detected unless the GW signal is caught at the time of its birth.

\reftitle{References}
\bibliographystyle{mdpi}
\bibliography{bibliography}




\end{document}